\documentclass[a4paper,11pt]{JHEP}

\usepackage{fancyheadings}
\usepackage[dvips]{graphicx}
\usepackage{ifthen}
\usepackage{amssymb}
\usepackage{epsfig}
\usepackage[errorshow]{tracefnt}

\newcommand\sss{\scriptscriptstyle}
\newcommand\scs{\scriptstyle}

\newcommand\benu{\begin{enumerate}}
\newcommand\eenu{\end{enumerate}}
\newcommand\bit{\begin{itemize}}
\newcommand\eit{\end{itemize}}

\newcommand{\be}{\begin{eqnarray}}
\newcommand{\ee}{\end{eqnarray}}
\newcommand{\bd}{\begin{displaymath}}
\newcommand{\ed}{\end{displaymath}}
\newcommand{\bq}{\begin{equation}}
\newcommand{\eq}{\end{equation}}

\newcommand\RR{{\mathbb R}}

\newcommand\ZZ{{\mathbb Z}}

\newcommand\wdg{\wedge}

\newcommand\h{\frac{1}{2}}

\newcommand\etal{{\it et al.} }
\newcommand\ie{{\it i.e.}}
\newcommand\eg{{\it e.g.}}

\title{Non-Commutative Open $(p,q)$-String Theories}

\author{Ulf Gran and Mikkel Nielsen\\Department of Theoretical Physics\\G\"oteborg University and Chalmers University of Technology\\SE-412 96 G\"oteborg, Sweden\\E-mail: \email{gran@fy.chalmers.se,mikkel@fy.chalmers.se}}

\abstract{
In this paper we make an SL(2,$\ZZ$)-covariant generalisation of the noncommutative theories, NCYM and NCOS on the D3-brane, and NCOS on the D5-brane in type IIB. Usually, the noncommutative theories are obtained by studying perturbative F-string theory, and the parameters governing the noncommutative theories are given by the open string data. The S-duality of NCYM and NCOS on the D3-brane has been seen by dualising the background, keeping the F-string theory under study fixed. We give an SL(2,$\ZZ$)-covariant generalisation of the open string data relevant when one instead studies perturbative $(p,q)$-string theory. The S-duality of NCYM and NCOS on the D3-brane is reproduced by instead keeping the background fixed and studying different $(p,q)$-string theories. We also obtain new noncommutative open $(p,q)$-string theories on the D3-brane and the D5-brane which are S-dual to ordinary NCOS.  The theories are studied using the supergravity duals of the D3-brane and the D5-brane, corresponding to a probe brane in the relevant background.}


\begin{document}
\section{Introduction}

Noncommutative theories on branes have been studied a lot in the recent couple of years. In particular, the theories on the D3-brane have received much attention. Noncommutative Yang-Mills, NCYM, was seen to arise in the zero slope limit of the gauge theory on a D$p$-brane in a magnetic $B$-field \cite{Seiberg:1999}. The effect of the $B$-field is to change the boundary conditions for open strings ending on the D$p$-brane. This, in turn, changes the propagator, from which one can read off the effective open string metric $G^{\mu\nu}$ and the noncommutativity parameter $\Theta^{\mu\nu}$, which lies in the magnetic directions. We also get an effective open string coupling $G_{{\sss {\mathrm O}}}$.
 The $\alpha'\rightarrow 0$ limit decouples the physics on the brane from the bulk. If we want to make sense of NCYM in this limit, it is the $G_{\mu\nu}, \Theta^{\mu\nu}$ and $G_{{\sss {\mathrm O}}}$ which should be kept fixed, and this yields the scaling of the closed string fields $g_{\mu\nu}, B_{\mu\nu}$ and $e^{\phi}$ in the limit. The $\alpha'\rightarrow 0$ limit decouples all excited string states, leading to a noncommutative field theory.

When one instead considers D-branes in a critical electric background, the situation is different, since the critical electric field leads to an effective finite open string tension $(\alpha'_{{\sss \mathrm{eff}}})^{-1}$, and the entire open string spectrum is kept. The closed string spectrum is still governed by $\alpha'$, so the result is again a decoupled theory on the brane, but this time a noncommutative theory of open strings, NCOS \cite{Gopakumar:2000:1,Seiberg:2000}. Furthermore, we now get space-time noncommutativity in the electric directions. It was also seen that NCOS arises in the strong coupling limit of NCYM and therefore the theories are S-dual.

The theories on the brane can also be studied via the dual description with one probe brane in a background of a stack with a large number of source branes.
The supergravity duals of the noncommutative theories have been described in \cite{Hashimoto:1999,Maldacena:1999} for NCYM and in \cite{Gopakumar:2000:1,Harmark:2000:1} for NCOS. In \cite{Berman:2000:2,Sundell:2000}, general electric and magnetic deformations of D-branes are described in a compact way, using the T-duality group. The open string data are calculated using the supergravity duals\footnote{This has also been done in \cite{Intriligator:1999,Lu:2000}. In \cite{Russo:2000:2} this was done for the open string metric and the noncommutativity parameter.}. The decoupling limit can be seen as a UV limit for the supergravity dual, \ie, the distance between the probe and source branes becomes large \cite{Li:1999,Chen:2000,Berman:2000:1,Berman:2000:2}.

In this paper we propose an SL(2,$\ZZ$)-covariant generalisation of the NCYM and NCOS theories. This is done by obtaining SL(2,$\ZZ$)-covariant open string data in which we insert the backgrounds of the supergravity duals, and the decoupling limit is obtained in the UV limit. The idea is as follows: Since type IIB string theory is supposed to have an exact nonperturbative SL(2,$\ZZ$) symmetry, this should manifest itself in the dynamics of the objects of this theory. NCYM and NCOS on the D3-brane have so far both been obtained from perturbation theory of fundamental strings ((1,0)-strings). The S-duality between the two theories on the D3-brane has been seen as the S-duality of the backgrounds, obtained by SL(2,$\ZZ$) transforming a magnetic $B$-field and an electric $C$-field to a magnetic $C$-field and an electric $B$-field. The electric $B$-field is responsible for the fact that (1,0)-strings are light in the NCOS case. Note that one is free to consider the effective theory on the brane, resulting by studying any open string that is allowed to end on the brane in a given background. Usually (1,0)-strings are used, since perturbation theory of these are well known. But one might as well study pertubation theory of $(p,q)$-strings to see what kind of theories we can get on the brane. In fact, instead of performing an SL(2,$\ZZ$)-transformation on the background with fixed (1,0)-strings as above, one may equivalently keep the background fixed and transform the strings\footnote{We are grateful to M. Cederwall and B.E.W. Nilsson for pointing this out to us.}. In this way the S-duality between NCOS and NCYM can be seen by transforming (1,0)-strings to (0,1)-strings in a fixed background with an electric $B$-field. This holds for vanishing axion. When the axion is rational, we reproduce the S-duality between NCOS and NCYM by transforming (1,0)-strings to $(p,q)$-strings, with $p-q\chi=0$, in a fixed background. 

That this procedure works is a consequence of the SL(2,$\ZZ$)-symmetry; transforming the entire system, \ie,  both the open strings we study and the background, yields an equivalent description. It is therefore clear that NCYM can be obtained either from studying (1,0)-strings in a background with a magnetic $B$-field, or equivalently from studying (0,1)-strings in a background with a magnetic $C$-field. Similarly, NCOS can be obtained either from studying (1,0)-strings in a background with an electric $B$-field or from studying (0,1)-strings in a background with an electric $C$-field. In fact, we get a whole SL(2,$\ZZ$)-orbit of equivalent descriptions of NCOS (and of NCYM), and we just choose a representative from this orbit, namely the one obtained from studying (1,0)-strings in a background of an electric $B$-field. 

When studying S-dual theories, the transformation on the background or the strings is, just as for the usual NCOS/NCYM case, performed before taking the decoupling limit, which is important since the theories on the brane should originate from string theory in a well defined way.

For arbitrary axion, we can start with NCOS, obtained from studying (1,0)-strings in the background with an electric $B$-field. We can then study $(p,q)$-strings in the same background and obtain an S-dual description even in the case where $p-q\chi\neq 0$. This is a new S-duality between ordinary NCOS and another NCOS, obtained from studying $(p,q)$-strings. We will show that the $(p,q)$-strings are light in the decoupling limit. The entire spectrum is therefore kept and therefore we get a noncommutative theory of open $(p,q)$-strings.

The above analysis can also be performed for the 5-branes in type IIB. Again we can keep the background fixed, we will choose a D5-brane in all possible 2-form configurations, ranging from rank 2 to rank 6. We get NCOS by studying (1,0)-strings in an electric $B$-field background and as before we can obtain an S-dual theory from studying $(p,q)$-strings\footnote{Note that we again have an SL(2,$\ZZ$)-orbit of equivalent theories obtained by transforming both the open strings and the background. NCOS can for example also be obtained by studying D-strings in a background of an electric $C$-field. Being equivalent to NCOS this theory is different from OD1 which is supposed to be S-dual to NCOS.}. In the rank 6 case, NCOS turns out to be S-dual to another NCOS, obtained from studying $(p,q)$-strings, with $p-q\chi=0$. For the lower rank cases, we also get several examples of NCOS being S-dual to $(p,q)$ NCOS for arbitrary values of the axion with $p-q\chi\neq 0$.

In section 2 we present the relevant supergravity duals and in section 3 we present the noncommutative open $(p,q)$-string theories. We end by discussing the results.

\section{The supergravity duals}

In this section we briefly discuss the 3-brane and 5-brane solutions, which in the near horizon limit become the supergravity duals of the noncommutative theories. As discussed in the previous section, we only need the (F,D3) and the D5-solutions.

As mentioned in the introduction, the noncommutative theories on the brane can be described by the open string data
\be\nonumber
&G^{\mu\nu}=\Big((g+B)_{{\sss S}}^{-1}\Big)^{\mu\nu}= \Big((g+B)^{-1}g(g-B)^{-1}\Big)^{\mu\nu}&\\
&\frac{\Theta^{\mu\nu}}{\alpha'}=\Big((g+B)^{-1}_{{\sss A}}\Big)^{\mu\nu}=-\Big((g+B)^{-1}B(g-B)^{-1}\Big)^{\mu\nu}&\\\nonumber
&G_{\mu\nu}=g_{\mu\nu}-B_{\mu\lambda}g^{\lambda\kappa}B_{\kappa\nu}&\\\nonumber
&G_{{\sss {\mathrm O}}}=e^{\phi}\sqrt{\frac{\mbox{{\rm det}}(g+B)}{\mbox{{\rm det}}g}}=e^{\phi}\Big(\frac{\mbox{{\rm det}}\,G}{\mbox{{\rm det}}\,g}\Big)^{\frac{1}{4}}&
\ee
where A, S refer to the antisymmetric and symmetric parts. Usually, the data are calculated in the decoupling limit of the brane in a fixed flat background. Alternatively, one can use the dual holographic picture and calculate the data using the fields of the supergravity solution. This is the approach we will take.

Our starting point are the type IIB supergravity solutions of Cederwall \etal \cite{Cederwall:1998,Cederwall:1999}. Our conventions for type IIB supergravity can be found appendix A.

\subsection{The D3-brane}
We start with the D3-brane solution \cite{Cederwall:1998}. This solution is the most general solution for D3-branes in $B$-fields, or equivalently, a bound state of (F,D1)-strings and D3-branes\footnote{Later, other solutions have appeared, \eg, \cite{Lu:1999:2}, but these can be obtained by a rescaling of the coordinates \cite{Gran:2001:1}.}, and it is parametrised by a complex anti-selfdual 2-form $F_{(2)}$. The radial dependence of the undeformed D3-brane solution is described by the harmonic function
\bq
\Delta=1+\frac{R^4}{r^4}
\eq
We can then define $\Delta_\pm=\Delta\pm\nu$, where $\nu$ describes the deformation. More precisely, $\nu=2\vert\mu\vert$, where $\mu={\scs \frac{1}{4}}\mbox{{\rm tr}}\,\big(F_{(2)}\big)^2$. The original deformed solution is an SL(2,$\ZZ$)-covariant description of all bound states of D3-branes and (F,D1)-strings. Picking a certain string amounts to choosing the scalar doublet ${\cal U}^r$, see appendix A for details. The solution for the (F,D3) bound state in the Einstein frame is \cite{Gran:2001:1}
\be\nonumber
ds^2&=&\Delta_+^{\frac{1}{4}}\Delta_-^{-\frac{3}{4}}\Big(-(dx^0)^2+(dx^1)^2\Big)+\Delta_+^{-\frac{3}{4}}\Delta_-^{\frac{1}{4}}\Big((dx^2)^2+(dx^3)^2\Big)+\Delta_+^{\frac{1}{4}}\Delta_-^{\frac{1}{4}}dy^2\\
{\cal U}^1&=&-{\scs \frac{1}{c}}\,\eta\,\Delta_+^{-\frac{1}{4}}\Delta_-^{\frac{1}{4}}\,,\qquad\qquad\ \
{\cal U}^2=i\,c\,\eta\,\Delta_+^{\frac{1}{4}}\Delta_-^{-\frac{1}{4}}\\
C_{1}&=&c\sqrt{2\nu}\,\Delta_-^{-1}\,dx^0\wdg dx^1\,,\quad
C_{2}=c^{-1}\sqrt{2\nu}\,\Delta_+^{-1}\,dx^2\wdg dx^3\\\nonumber
e^\phi&=&c^{2}\sqrt{{\scs \frac{\Delta_+}{\Delta_-}}}\,,\qquad\qquad\qquad\quad\ \ \chi=0
\ee
where $c$ is an arbitrary real constant (actually $c^2$ is the undeformed asymptotic dilaton) and $\eta=\mu/\vert\mu\vert$. As explained in the appendix, the 2-forms, usually written as $B$ and $C$, are now collected in a doublet of 2- forms as $C_{1}$ and $C_{2}$. Non-vanishing axion can be obtained by doing an SL(2,$\RR$)-transformation which in general changes the string charges $(p^1,p^2)=(p,0)$ to an arbitrary
doublet $(p,q)$
\bd
\left(\begin{array}{c}p\\q\end{array}\right)=\left(\begin{array}{cc}1&p\tilde{p}\\q/p&p
\tilde{q}\end{array}\right)\left(\begin{array}{c}p\\0\end{array}\right)
\ed
where $\tilde{p},\tilde{q}$ are real numbers fulfilling
$p\tilde{q}-q\tilde{p}=1$. So the case $q$=0 yields
$\tilde{q}$=$1/p$, and keeping $\tilde{p}\neq 0$ gives the
solution with general axion. The scalar doublet transforms in the
same way as the charge doublet
\bd
\left(\begin{array}{c}\tilde{{\cal U}}^1\\\tilde{{\cal
U}}^2\end{array}\right)= \left(\begin{array}{c}{\cal
U}^1+p\tilde{p}\,{\cal U}^2\\{\cal U}^2\end{array}\right)
\ed
and we immediately see that the axion is given by
\bd
\chi=p\tilde{p}
\ed
The doublet of 2-forms becomes
\be\label{cchi}
C_{1}&=&c\sqrt{2\nu}\,\Delta_-^{-1}\,dx^0\wdg dx^1\\\nonumber
C_{2}&=&-\chi\,c\sqrt{2\nu}\,\Delta_-^{-1}\,dx^0\wdg dx^1+c^{-1} \sqrt{2\nu}\,\Delta_+^{-1}\,dx^2\wdg dx^3 \ee

\subsection{The D5-brane}
Now turn to the D5-brane solution with arbitrary rank of the
$B$-field. The solution is by construction half-supersymmetric \cite{Cederwall:1999} and can be seen as a bound state of a D5-brane with all possible lower dimensional branes, depending on the rank of the $B$-field. In its original form \cite{Cederwall:1999}, the
solution looks quite complicated, but it can be simplified by
using a particular basis, see \cite{Gran:2001:1} for details. The
radial dependence is now given by the harmonic function \bq
\Delta=1+\frac{R^2}{r^2} \eq Then we can define the deformed
harmonic functions \bq
\Delta_{\sss{\pm\pm\pm}}=\Delta\pm\nu_1\pm\nu_2\pm\nu_3 \eq where
$\nu_i$ corresponds to ${\scs \frac{9}{8}}\,\tilde{\nu}_i^2$ in
\cite{Cederwall:1999}. Then the solution in the  Einstein frame
takes the following form 
\be\label{5metric}\nonumber\label{d5}
&&ds^2=\Delta_{\sss{---}}^{-3/4}(\Delta_{\sss{++-}}\Delta_{\sss{+-+}})^{1/4}
\big(-dx_0^2+dx_1^2\big)+\Delta_{\sss{++-}}^{-3/4}(\Delta_{\sss{---}}
\Delta_{\sss{+-+}})^{1/4}\big(dx_2^2+dx_3^2\big)\\\nonumber
&&\hspace{1cm}+\
\Delta_{\sss{+-+}}^{-3/4}(\Delta_{\sss{---}}\Delta_{\sss{++-}})^{1/4}
\big(dx_4^2+dx_5^2\big)+(\Delta_{\sss{---}}\Delta_{\sss{++-}}\Delta_{\sss{+-+}})^{1/4}dy^2\\\nonumber
&&(C_{1})_{01}=k^{-1}\sqrt{2\nu_1}\,\Delta_{\sss{---}}^{-1}\,,\qquad
(C_{2})_{01}=-2k\sqrt{\nu_2\nu_3}\Delta_{{\sss
---}}^{-1}-k^{-1}\tilde{q}\sqrt{2\nu_1}\,\Delta_{\sss{---}}^{-1}\\\nonumber
&&(C_{1})_{23}=k^{-1}\sqrt{2\nu_2}\,\Delta_{\sss{++-}}^{-1}\,,\qquad
(C_{2})_{23}=2k\sqrt{\nu_1\nu_3}\Delta_{{\sss
---}}^{-1}-k^{-1}\tilde{q}\sqrt{2\nu_2}\,\Delta_{\sss{++-}}^{-1}\\\nonumber
&&(C_{1})_{45}=k^{-1}\sqrt{2\nu_3}\,\Delta_{\sss{+-+}}^{-1}\,,\qquad (C_{2})_{45}=2k\sqrt{\nu_1\nu_2}\Delta_{{\sss +-+}}^{-1}-k^{-1}\tilde{q}\sqrt{2\nu_3}\,\Delta_{\sss{+-+}}^{-1}\\
&&e^{\phi}=k^{-2}\Big(\Delta_{{\sss ---}}\Delta_{{\sss ++-}}\Delta_{{\sss +-+}}\Big)^{-\frac{1}{2}}\Delta_{{\sss +--}}\,,\qquad \chi=\tilde{q}-k^2\sqrt{8\nu_1\nu_2\nu_3}\Delta_{{\sss +--}}^{-1}
\ee
where $k$ is a real constant related to the asymptotic scalars and $\tilde{q}$ is a real parameter. For general 5-brane charges $(p_1,p_2)=(p,q)$,  we instead have a real doublet $(\tilde{p}_1,\tilde{p}_2)=(\tilde{p},\tilde{q})$, fulfilling $\epsilon^{rs}p_r\tilde{p}_s=1$. The D5 solution is equivalent to the one obtained without the RR-fields in \cite{Alishahiha:1999}. The rank 2 case was first found in \cite{Lu:1999:3}.

We are going to use the solutions above for the generalised open $(p,q)$-string theories in the next section.

\section{Open $(p,q)$-string theories}

As mentioned, NCYM and NCOS on the D3-brane can be described in a setup where fundamental strings are attached to a probe brane, sitting in a background magnetic and electric $B$-field, respectively. In particular, the important quantities, the open string metric, the noncommutativity parameter and  the open string coupling are usually derived from perturbation theory of the fundamental string. The S-duality of NCYM and NCOS has so far been seen as the S-duality of the backgrounds, but alternatively one can get an S-dual description by using the dual strings, \ie, D-strings to obtain the theory on the probe brane as discussed in the introduction. This holds for vanishing axion. For non-vanishing axion, we are lead to study open $(p,q)$-strings ending on the D3-brane in the same background when searching for an S-dual theory. On the type IIB 5-branes, we get a similar picture. NCOS is obtained by studying an (1,0)-strings in a background of an electric $B$-field\footnote{We do not consider the little strings on the 5-branes. Including these might alter some of the  conclusions reached.}, and we look for an S-dual theory by studying $(p,q)$-strings in the same background.

Doing an SL(2,$\ZZ$) transformation on just the strings ending on the probe is equivalent to just transforming the background. Studying $(p,q)$-string theory in a certain background is therefore equivalent to studying (1,0)-string theory in a transformed background. The latter approach was persued in \cite{Russo:2000,Cai:2000} for the D3-brane (and in \cite{Lu:2000}, but they only consider the usual S-duality, which is just one specific SL(2,$\ZZ$)-element, on the general background). Here we will concentrate on the first approach and the general SL(2,$\ZZ$)-duality. It should be clear that the noncommutative theories of open $(p,q)$-string theories we obtain on the D3- and the D5-branes are equivalent to NCOS theories in a transformed background, but it is important to keep in mind that the new theories will be S-dual to ordinary NCOS instead of being equivalent to this. One of our motivations for the present approach is that it is simpler to transform a charge doublet than transforming the whole background.

We want an SL(2,$\ZZ$)-covariant description of the open string data relevant for the noncommutative theories. Thus, our starting point is the Einstein metric, the doublet of 2-forms and  the scalar doublet instead of the string metric, the $B$-field, the dilaton and the axion. As should be clear from the discussion above, when considering the 2-forms, it is important what kind of strings we are using for our description. To be precise, it is the angle between the open string charges and the 2-forms that matters. In the open string data, we should therefore replace $B$ with $p^r C_r$, where $(p^1,p^2)=(p,q)$ are the charges of the open strings ending on the probe brane. The charges with indices downstairs are $p_r=\epsilon_{rs}p^s$, and therefore\footnote{We use the convention, $\epsilon^{12}=1$ and $\epsilon_{12}=-1$.} $(p_1,p_2)=(-q,p)$. In the Im(${\cal U}^2$)=0 gauge, the $(p,q)$-string tension in the Einstein frame is, in units of ${\scs \frac{1}{\alpha'}}$ \cite{Schwarz:1995,Cederwall:1997:1},
\bq
\vert{\cal U}^r p_r\vert=\sqrt{e^{\phi}(p-q\chi)^2+q^2 e^{-\phi}}
\eq
And we get the SL(2,$\ZZ$)-covariance by replacing $e^{\phi/2}$ everywhere with $\vert{\cal U}^r p_r\vert$. Then the $(p,q)$ open string data become
\be\nonumber
&G^{\mu\nu}&\!=\!{\scs \frac{1}{\vert{\cal U}^r p_r\vert}}\Big(\big(g^{{\sss \mathrm{E}}}\!+\!{\scs \frac{p^s C_{s}}{\vert{\cal U}^t p_t\vert}}\big)_{{\sss \mathrm{S}}}^{-1}\Big)^{\mu\nu}\!=\!{\scs \frac{1}{\vert{\cal U}^r p_r\vert}}\Big(\big(g^{{\sss \mathrm{E}}}\!+\!{\scs \frac{p^s C_{s}}{\vert{\cal U}^t p_t\vert}}\big)^{-1}g^{{\sss \mathrm{E}}}\big(g^{{\sss \mathrm{E}}}\!-\!{\scs \frac{p^m C_{m}}{\vert{\cal U}^n p_n\vert}}\big)^{-1}\Big)^{\mu\nu}\\\nonumber
&\frac{\Theta^{\mu\nu}}{\alpha'}&\!=\!{\scs \frac{1}{\vert{\cal U}^r p_r\vert}}\Big(\big(g^{{\sss \mathrm{E}}}\!+\!{\scs \frac{p^s C_{s}}{\vert{\cal U}^t p_t\vert}}\big)_{{\sss \mathrm{A}}}^{-1}\Big)^{\mu\nu}\!=\!-{\scs \frac{1}{\vert{\cal U}^r p_r\vert}}\Big(\big(g^{{\sss \mathrm{E}}}\!+\!{\scs \frac{p^s C_{s}}{\vert{\cal U}^t p_t\vert}}\big)^{-1}{\scs \frac{p^k C_{k}}{\vert{\cal U}^l p_l\vert}}\big(g^{{\sss \mathrm{E}}}\!-\!{\scs \frac{p^m C_{m}}{\vert{\cal U}^n p_n\vert}}\big)^{-1}\Big)^{\mu\nu}\\
&G_{\mu\nu}&\!=\!\vert{\cal U}^r p_r\vert\Big(g^{{\sss \mathrm{E}}}_{\mu\nu}-\,{\scs \frac{(p^s C_{s})^2_{\mu\nu}}{\vert{\cal U}^t p_t\vert^2}}\Big)\,,\quad G_{\sss \mathrm{O}}\!=\!\vert{\cal U}^r p_r\vert^{\frac{7-p}{4}}\Big(\frac{\mathrm{det}\,G}{\mathrm{det}\,g^{{\sss \mathrm{E}}}}\Big)^{\frac{1}{4}}
\ee
It is easy to see that the above formulae reduce to the usual when $(p,q)=(1,0)$. It is also important to remember that the charges $p^r$ of the open strings ending on the probe brane are independent of the background consisting of the metric $g^{{\sss \mathrm{E}}}$, the 2-forms $C_{r}$ and the scalars ${\cal U}^r$. The open string data are therefore constructed from two separate sets of SL(2,$\ZZ$)-covariant quantities. Since an SL(2,$\ZZ$) duality transformation corresponds to transforming just one of these sets, the open string data are not invariant under the duality transformation but transform covariantly. Furthermore, the fact that one can SL(2,$\ZZ$)-transform either the open strings or the background, yielding the same theory on the probe, is manifest in the formulas above. By transforming both the background as well as the strings ending on the probe we get an SL(2,$\ZZ$)-orbit of equivalent theories with the same open string data.

\subsection{The D3-brane}

The supergravity duals are obtained in certain scaling limits,
corresponding to $\alpha'\rightarrow 0$. If we use the
limits obtained in \cite{Berman:2000:2} and translate them to our coordinates, the result for the D3-brane is that the coordinate
scalings are the same in both the electric and the magnetic near
horizon limits. 
\bq
\hat{x}=\ell\,\frac{x}{\sqrt{\alpha'}}\,,\quad
u=\ell\,\frac{r}{\sqrt{\alpha'}}\,,\quad c\qquad \mathrm{fixed}
\eq
where $\ell$ is a fixed length scale. So, whereas one usually has different
scalings of the radial coordinate, \ie, ${\sss \frac{r}{\alpha'}}$ should be
fixed in the NCYM case and ${\sss \frac{r}{\sqrt{\alpha'}}}$ should be fixed in
the NCOS case, we have the same scaling in both cases. There is no freedom in
the scaling above. It can be derived by demanding that $ds^2/\alpha'$ are fixed
as $\alpha'$ goes to zero, which also implies that $C_{r}/\alpha'$ are
fixed in this limit, yielding a finite supergravity action. Since $ds^2/\alpha'$
is SL(2,$\ZZ$) invariant ($ds^2$ is evaluated using the Einstein metric), the
above scalings are obtained independently of what background we choose (if we
were to SL(2,$\ZZ$)-transform transform the background instead of the string
theory under study). 
There is thus a unique scaling limit for both NCYM, NCOS and open
$(p,q)$-string theories. 

From the supergravity point of view, the decoupling limit for the theories on the brane is obtained in three steps. We first go to critical field strength, enabling the
decoupling of closed strings. We then take the $\alpha'\rightarrow 0$ limit, which decouples the massless closed strings. We finally take the UV limit, which decouples the massive closed strings and in the NCYM case also decouples the massive open
strings. In the UV limit, the open strings ending on the probe brane will be "sucked" into the brane since the open string tension in the transverse directions diverges whereas the effective open string tension will be finite on the brane in this limit. The resulting decoupled theories are therefore obtained from the string/supergravity setup in a well defined way from a unique decoupling limit. 

For general charges we can then calculate the data, using fixed coordinates and starting with the electric background with constant axion from the previous section
\be\nonumber\label{pqdata}
&\frac{G_{\mu\nu}}{\alpha'}&=\frac{1}{\ell^2}\,c^{-1}(q^2+c^4(p-q\chi)^2)(q^2\Delta_-+c^4(p-q\chi)^2\Delta_+)^{-\frac{1}{2}}\eta_{\mu\nu}\\
&\Theta^{01}&=-\ell^2\sqrt{2\nu}\,c^3(p-q\chi)(q^2+c^4(p-q\chi)^2)^{-1}\\\nonumber
&\Theta^{23}&=-\ell^2\sqrt{2\nu}\,c\,
q\,(q^2+c^4(p-q\chi)^2)^{-1}\\\nonumber &G_{{\sss
\mathrm{O}}}&=c^{-2}(q^2+c^4(p-q\chi)^2) 
\ee 
As in \cite{Berman:2000:2}, $\Theta$ and $G_{{\sss \mathrm{O}}}$ are
$r$-independent. In \cite{Berman:2000:2} infinite magnetic
deformation parameter is obtained in the $\alpha'\rightarrow 0$
limit and the critical electrical field is obtained in the limit
${\scs \frac{\theta}{\alpha'}}\rightarrow 1$. In our case both the
limits corresponds to $\nu\rightarrow 1$. In this sense the
electric and magnetic deformations are described in a unified way
with our coordinates. In particular, $\nu\rightarrow 1$ can be
interpreted as a critical field limit independently of the field
being electric or magnetic. The UV limit corresponds to large $u$
and therefore we can write the harmonic function as \bq
\Delta=1+\frac{\hat{R}^4}{u^4}\sim1+\epsilon \eq Using the
critical field limit, the result for the deformed harmonic
functions is \bq \Delta_{-}\sim \epsilon\,,\quad\Delta_+\sim 2 \eq
The usual S-duality between electric and magnetic backgrounds now
holds for the $(p,q)=(1,0)$ and the (0,1) strings ending on the probe but with a
background with vanishing axion, since then we get electric and
magnetic noncommutativity parameter, respectively. The open string
data for the (1,0)-string are 
\be 
&{\scs
\frac{G_{\mu\nu}}{\alpha'}}={\scs
\frac{c}{\sqrt{2}\ell^2}}\,\eta_{\mu\nu}\,,\qquad
\Theta^{01}=-\ell^2\sqrt{2}\,c^{-1}\label{NCOSdata}\\\nonumber &G_{{\sss
\mathrm{O}}}=c^2\,,\qquad \Theta^{23}=0& \ee and for the
(0,1)-string we have \be &{\scs \frac{G_{\mu\nu}}{\alpha'}}={\scs
\frac{1}{\ell^2}}\,c^{-1}\epsilon^{-{\scs
\frac{1}{2}}}\eta_{\mu\nu}\,,\qquad\Theta^{23}=-\ell^2\sqrt{2}\,c&\\\nonumber
&G_{{\sss \mathrm{O}}}=c^{-2}\,,\qquad \Theta^{01}=0& 
\ee
 and thus the open string coupling for the NCOS is the inverse of that of the NCYM and the result is therefore a strong/weak coupling duality. We have thus reproduced the usual S-duality between NCOS and NCYM. This is not a surprise, since it is clear from our SL(2,$\ZZ$)-covariant open string data that our approach of dualising the open strings ending on the probe brane is equivalent to the usual approach of just dualising the background.

If we instead consider the charges (1,0) and non-vanishing axion
we still get NCOS, $\Theta^{01}\neq 0$ and $\Theta^{23}=0$, as we
see from (\ref{pqdata}) with $(p,q)=(1,0)$. Now do an
SL(2,$\ZZ$)-transformation to new charges $(p,q)$, with $q\neq 0$.
The open string data are still given by (\ref{pqdata}), but now
with the general charges. Then we get a NCYM if $p-q\chi=0$, which
can only be achieved if the axion is rational. The open string
coupling for the NCOS is $G_{{\sss \mathrm{O}}}=c^2$, and for the
NCYM it is $G_{{\sss \mathrm{O}}}=c^{-2} q^2$, so if we let $c$ go
to zero or infinity we see that a strongly coupled NCOS
corresponds to a weakly coupled NCYM and vice versa, \ie, the
theories are S-dual. This was first derived in \cite{Russo:2000}
(and it was also discussed in \cite{Lu:2000:2} but only for the
theories on the brane and not for the supergravity duals). Again, by just transforming the open strings ending on the probe brane, we have reproduced the known result that NCOS and NCYM are S-dual for a rational axion, which had previously been obtained by just transforming the background .

For arbitrary axion and general charges for the strings on the probe, we always get NCOS theories if $p-q\chi\neq 0$, since then $\Theta^{01}\neq 0$. This case certainly includes all irrational values of the axion and arbitrary charges for the strings, and almost all cases of rational values of the axion with arbitrary string charges (the obvious exception being the one set of charges fulfilling $p-q\chi=0$). When $q\neq 0$ we also get $\Theta^{23}\neq 0$. In fixed coordinates, the open string data are
\be\nonumber
&\frac{G_{\mu\nu}}{\alpha'}&={\scs \frac{1}{\sqrt{2}\ell^2}}\,c^{-3}(q^2+c^4(p-q\chi)^2)\vert p-q\chi\vert^{-1}\eta_{\mu\nu}\\
&\Theta^{01}&=-\ell^2\sqrt{2}\,c^3(p-q\chi)(q^2+c^4(p-q\chi)^2)^{-1}\\\nonumber
&\Theta^{23}&=-\ell^2\sqrt{2}\,c\, q\,(q^2+c^4(p-q\chi)^2)^{-1}\\\nonumber
&G_{{\sss \mathrm{O}}}&=c^{-2}q^2+c^2(p-q\chi)^2
\ee
But two such theories can in general never be S-dual, as noted in \cite{Lu:2000:2}, since the coupling has the form $G_{{\sss \mathrm{O}}}=c^{-2}q^2+c^2(p-q\chi)^2$, and the second term is always non-vanishing. Consider for instance charges (1,0) and $(p,q)$. Even though the couplings go to zero and infinity, respectively, when $c\rightarrow 0$, both go to infinity for $c\rightarrow\infty$. There is, however, a special case where an S-duality can be obtained. For arbitrary axion and large $c$, we can find charges $(p,q)$ such that ${\scs \frac{p}{q}}-\chi\sim c^{-\beta}$, where $\beta>1$. This statement just corresponds to the fact that for any real number, we can find a rational number as close to the real number as we want. Then the coupling scales with a negative power of $c$ and the result is a strong/weak coupling duality between two NCOS theories, the usual NCOS and a new NCOS obtained from open $(p,q)$-strings. Note that the new NCOS theory is not equivalent to ordinary NCOS, since it is S-dual to this theory. Furthermore, the theory will in general have both space-time and space-space noncommutativity. For $\beta=2$ we get by simply replacing ${\scs \frac{p}{q}}-\chi$ with $c^{-2}$
\be\nonumber
&\frac{G_{\mu\nu}}{\alpha'}&\simeq{\scs \frac{\sqrt{2}}{\ell^2}}\,c^{-1}q\eta_{\mu\nu}\\
&\Theta^{01}&\simeq-{\scs \frac{\ell^2}{\sqrt{2}}}\,c\,q^{-2}\\\nonumber
&\Theta^{23}&\simeq-{\scs \frac{\ell^2}{\sqrt{2}}}\,c\,q^{-1}\\\nonumber
&G_{{\sss \mathrm{O}}}&\simeq 2 c^{-2}q^2
\ee
We would like to stress that this S-duality is somewhat different than the usual, for the following reason. Once we have picked charges $(p,q)$ such that ${\scs \frac{p}{q}}-\chi\sim c^{-2}$, we want to keep these charges to have a well-defined $(p,q)$ NCOS. If one then increases the coupling of (1,0) NCOS, one must change the axion such that the required scaling is still obeyed. The S-duality thus involves a change in the parameters $c$ and $\chi$, \ie, both the scalars of the undeformed solution. It also follows from the above that even in the case of a rational axion we can get an S-dual noncommutative open $(p,q)$-string theory instead of NCYM; ${\sss \frac{p}{q}}$ just has to be close to $\chi$ for large $c$, and once we have picked such a set of charges, we again have to tune the axion when $c$ is increased. In this sense we actually get a $(p,q)$ NCOS in a generic case and we only obtain NCYM in certain special cases.

In general it is seen that ${\scs \frac{G_{\mu\nu}}{\alpha'}}$ is finite in the UV when $p-q\chi\neq 0$. When $p-q\chi=0$, this quantity diverges in the UV.
The effective open string tension can be read off from the open string metric
\bq
\frac{1}{\alpha'_{{\sss \mathrm{eff}}}}=-\frac{G_{00}}{\alpha'}
\eq
which can be derived by considering the rest mass of an open string state in a curved background.  From the behaviour of the open string metric, we see that, as expected, the effective open string tension is finite for the NCOS-theories but diverges for NCYM, corresponding to the fact that we only have light strings for NCOS theories. Note that the formula above holds for whatever kind of strings we are using. In particular, when $p-q\chi\neq 0$, the open $(p,q)$-strings are light, which they should be for a noncommutative open $(p,q)$-string theory. Hence we do not need electric components of both the 2-forms to get light $(p,q)$-strings, since this would correspond to ordinary NCOS (being another element of the SL(2,$\ZZ$)-orbit of theories equivalent to NCOS). We can also calculate the tension by considering the string end points as a dipole and then taking into account the compensating force from the electric field. The usual flat space tension formula ${\scs \frac{1}{\alpha'_{\mathrm{eff}}}}={\scs \frac{1}{\alpha'}}-\epsilon^{01} F_{01}$ can also be written in a curved background as\footnote{This tension formula can also be generalised to open D-branes and membranes in curved backgrounds, related to OD$p$ and OM theories.}
\begin{equation}
\frac{1}{\alpha'_{\mathrm{eff}}}=\frac{\sqrt{-\mathrm{det}g_s}}{\alpha'}-\frac{\epsilon^{01}B_{01}}{\alpha'}
\end{equation}
where $g_s$ is the induced metric on the world sheet of the string. The first part diverges in the UV, but in the NCOS case we get a cancellation from the the second part. This formula can be generalised to an SL(2,$\ZZ$)-covariant form as follows
\begin{equation}
\frac{1}{\alpha'_{\mathrm{eff}}}=\vert{\cal U}^rp_r\vert\frac{\sqrt{-\mathrm{det}g_{{\sss \mathrm{E}}}}}{\alpha'}-\frac{\epsilon^{01}(C_{(2)r}p^r)_{01}}{\alpha'}
\end{equation}
With our fixed background, this expression transforms covariantly when we transform the open string charges $p^r$. Inserting the (F,D3) background and writing $\Delta_-=\epsilon$ and $\Delta_+=2+\epsilon$ the effective tension for $(p,q)$-strings, with $p-q\chi\neq 0$, becomes
\begin{equation}
\frac{1}{\alpha'_{\mathrm{eff}}}\simeq\frac{c}{\sqrt{2}\ell^2}(p-q\chi)
\end{equation}
And therefore the existence of light $(p,q)$-strings can also be seen from this formula. Note that both in the usual NCOS case with (1,0)-strings and in the S-dual NCOS case with ${\scs \frac{p}{q}}-\chi\simeq c^{-2}$ the effective tension obtained above just differs by the one read off from the open string metric by a factor of two. When $p-q\chi=0$, corresponding to a NCYM, we see from (\ref{cchi}) that $p^rC_{(2)r}=0$ and therefore there is no compensating effective electric field, and the result is a diverging tension as expected.

We have seen that the usual NCOS theory can be obtained from perturbative (1,0)-strings, and we then study the dual by just transforming the charges of the strings ending on the probe. For arbitrary axion, we can find charges such that NCOS is S-dual to a theory, obtained from studying $(p,q)$-strings. This theory has space-time noncommutativity and contains light $(p,q)$-strings, and it is therefore a  noncommutative theory of open $(p,q)$-strings. For the particular case $p-q\chi=0$, we reproduce the S-duality between (1,0) NCOS and NCYM, obtained from $(p,q)$-strings.

\subsection{$(p,q)$ 5-branes}

We can now do the same analysis for the $(p,q)$ 5-branes, using
the general rank solution from the previous section. As mentioned, we
only need to consider the D5 background, since the
SL(2,$\ZZ$)-duals are obtained by just transforming the charges of the
strings on the probe.  We consider each rank separately, beginning
with the highest rank.

\subsubsection{Rank 6}
In the near horizon limit, the following quantities are fixed
\bq
\hat{x}=\ell\,\frac{x}{\sqrt{\alpha'}}\,,\quad
u=\ell\,\frac{r}{\sqrt{\alpha'}}\,,\quad k\qquad \mathrm{fixed}
\eq
where $\ell$ is a fixed length scale. Just as for the D3-brane this scaling limit is unique, and would hold even if we transform the background. In the rank 6 case the critical field is obtained for $\nu_1+\nu_2+\nu_3\rightarrow 1$. The values of each of the parameters are undetermined, but each must be finite in the rank 6 case. One might worry whether it is allowed to consider SL(2,$\ZZ$)-dual strings, since D-strings cannot end on D5-branes. However, the background solution correspond to an (F,D1,D3,D3,D5) bound state, which can be seen from the 2- and 4-form potentials. We have electric NS-NS and RR 2-forms, yielding the strings in the bound state. From the 4-forms \cite{Gran:2001:1} we see that we get a D3-brane along $x^0,x^1,x^2,x^3$ and another along $x^0,x^1,x^4,x^5$, and the D-strings can end on these. As mentioned, the solution is half-supersymmetric by construction. One might be surprised by the presence of the D-string in the bound state, but this is just due to the presence of the electric RR 2-form for non-vanishing constant part of the axion, see (\ref{d5}). The D-string is therefore also present in the lower rank cases $\nu_2=0$ and/or $\nu_3=0$ as long as $\tilde{q}\neq 0$. However, it is of course not possible to obtain the one quarter supersymmetric (D1,D5), since the only way to get rid of the F-string is to put $\nu_1=0$, and by doing that, the D-string also disappears, which is easily seen from (\ref{d5}). Now define
\bq
f=\Big({\cal Q}\,\Delta_{{\sss +--}}+2\sqrt{2}k^2 q\sqrt{\nu_2\nu_3}\Big)^2+k^4 q^2\Delta_{{\sss ---}}\Delta_{{\sss +-+}}\Delta_{{\sss ++-}}
\eq
with
\bq
{\cal Q}=p-q\tilde{q}
\eq
Then the open string data for arbitrary charges of the open strings become
\be\nonumber\label{rank6}
\Theta^{01}&=&\ell^2k\,(2 k^2 q\sqrt{\nu_2\nu_3}-\sqrt{2\nu_1}{\cal Q})\Delta_{{\sss ---}}\Delta_{{\sss +--}}\Big(f-\Delta_{{\sss +--}}(2 k^2 q\sqrt{\nu_2\nu_3}-\sqrt{2\nu_1}{\cal Q})^2\Big)\\\nonumber
\Theta^{23}&=&-\ell^2k\,(2 k^2 q\sqrt{\nu_1\nu_3}\!+\!\sqrt{2\nu_2}{\cal Q})\Delta_{{\sss ++-}}\Delta_{{\sss +--}}\Big(f\!+\!\Delta_{{\sss +--}}(2 k^2 q\sqrt{\nu_1\nu_3}\!+\!\sqrt{2\nu_2}{\cal Q})^2\Big)\\\nonumber
\Theta^{45}&=&-\ell^2k\,(2 k^2 q\sqrt{\nu_1\nu_2}\!+\!\sqrt{2\nu_3}{\cal Q})\Delta_{{\sss +-+}}\Delta_{{\sss +--}}\Big(f\!+\!\Delta_{{\sss +--}}(2 k^2 q\sqrt{\nu_1\nu_2}\!+\!\sqrt{2\nu_3}{\cal Q})^2\Big)\\\nonumber
{\scs \frac{G_{11}}{\alpha'}}&=&{\scs \frac{1}{\ell^2}}\,k^{-1}\,\Delta_{{\sss ---}}^{-1}\Delta_{{\sss +--}}^{-\frac{1}{2}}f^{-\frac{1}{2}}\Big(f-\Delta_{{\sss +--}}(2 k^2 q \sqrt{\nu_2\nu_3}-\sqrt{2\nu_1}{\cal Q})^2\Big)\\\nonumber
{\scs \frac{G_{33}}{\alpha'}}&=&{\scs \frac{1}{\ell^2}}\,k^{-1}\,\Delta_{{\sss ++-}}^{-1}\Delta_{{\sss +--}}^{-\frac{1}{2}}f^{-\frac{1}{2}}\Big(f+\Delta_{{\sss +--}}(2 k^2 q \sqrt{\nu_1\nu_3}+\sqrt{2\nu_2}{\cal Q})^2\Big)\\
{\scs \frac{G_{55}}{\alpha'}}&=&{\scs \frac{1}{\ell^2}}\,k^{-1}\,\Delta_{{\sss +-+}}^{-1}\Delta_{{\sss +--}}^{-\frac{1}{2}}f^{-\frac{1}{2}}\Big(f+\Delta_{{\sss +--}}(2 k^2 q \sqrt{\nu_1\nu_2}+\sqrt{2\nu_3}{\cal Q})^2\Big)\\\nonumber
G_{{\sss \mathrm{O}}}&=&k^{-2}(\Delta_{{\sss ---}}\Delta_{{\sss +-+}}\Delta_{{\sss ++-}})^{-\frac{1}{2}}\Delta_{{\sss +--}}^{-1}f^{-\frac{1}{2}}\Big\vert f-\Delta_{{\sss +--}}(2 k^2 q\sqrt{\nu_2\nu_3}-\sqrt{2\nu_1}{\cal Q})^2\Big\vert^{\frac{1}{2}}\\\nonumber
&&\!\Big(f\!+\!\Delta_{{\sss +--}}(2 k^2 q\sqrt{\nu_1\nu_3}\!+\!\sqrt{2\nu_2}{\cal Q})^2\Big)^{\frac{1}{2}}\Big(f\!+\!\Delta_{{\sss +--}}(2 k^2 q\sqrt{\nu_1\nu_2}\!+\!\sqrt{2\nu_3}{\cal Q})^2\Big)^{\frac{1}{2}}
\ee
where the open string metric consists of three two-dimensional blocks. In contrast to the D3-brane case, the $\Theta$'s and $G_{{\sss \mathrm{O}}}$ are now $r$-dependent.
We now want to analyse these quantities in the UV limit which we define as follows
\bq
\Delta=1+\frac{\hat{R}^2}{u^2}\sim 1+\epsilon
\eq
Taking the critical field limit into account, the harmonic functions of the 5-brane become
\be
&&\Delta_{{\sss ---}}\sim \epsilon\,,\qquad\Delta_{{\sss ++-}}\sim 1+\nu_1+\nu_2-\nu_3\equiv a\\\nonumber
&&\Delta_{{\sss +-+}}\sim1+\nu_1-\nu_2+\nu_3\equiv b\,,\qquad \Delta_{{\sss +--}}\sim1+\nu_1-\nu_2-\nu_3\equiv c
\ee
and $a$, $b$ and $c$ are finite constants in the genuine rank 6 case. The strategy is now to insert different choices of charges of the strings on the probe and see what kind of theory emerges. We start with the charges (1,0), corresponding to an F-string, and then we consider $(p,q)$-strings to see if we can get an  S-dual description. In the (1,0) case we get
\be\nonumber
&\Theta^{01}=-\ell^2k\,\sqrt{2\nu_1}\,,\quad\Theta^{23}=-\ell^2k\,\sqrt{2\nu_2}\,,\quad\Theta^{45}=-\ell^2k\,\sqrt{2\nu_3}&\\
&\frac{G_{\mu\nu}}{\alpha'}=\frac{1}{\ell^2}\,k^{-1}c^{-\frac{1}{2}}\eta_{\mu\nu}\,,\qquad G_{{\sss \mathrm{O}}}=k^{-2}c^{-\frac{1}{2}}&
\ee
As expected we get an NCOS, but with both space-time and space-space noncommutativity. The effective open string tension can be read off from the effective open string metric
\bq
\frac{1}{\alpha'_{{\sss \mathrm{eff}}}}=-\frac{G_{00}}{\alpha'}=\frac{1}{\ell^2}\,k^{-1}c^{-\frac{1}{2}}
\eq
and the result is finite as it should be for an NCOS. The effective open string coupling is also finite and given by the undeformed coupling specified by the asymptotic dilaton.

Now consider general string charges, which is obtained by an SL(2,$\ZZ$)-transforma-\\tion on the F-string. Starting with the case ${\cal Q}=0$, which can be obtained for a rational value of the constant part of the axion, the result is
\be\nonumber
&\Theta^{01}=2\ell^2k^{-1}q^{-1}c\,\sqrt{\nu_2\nu_3}\,(ab-4\nu_2\nu_3)^{-1}\,,\quad \Theta^{23}=-\frac{1}{2}\,\ell^2k^{-1}q^{-1}c\,\sqrt{\nu_1\nu_3}&\\\nonumber
&\Theta^{45}=-\frac{1}{2}\,\ell^2k^{-1}q^{-1}c\,\sqrt{\nu_1\nu_2}\,,\quad
\frac{G_{11}}{\alpha'}=\frac{1}{2\ell^2}\,c^{-\frac{1}{2}}k q(2\nu_1\nu_2\nu_3)^{-\frac{1}{2}}\,(ab-4\nu_2\nu_3)&\\
&\frac{G_{33}}{\alpha'}=\frac{2}{\ell^2}\,c^{-\frac{1}{2}}k q\,(2\nu_1\nu_2\nu_3)^{-\frac{1}{2}}\nu_1\nu_3\,,\quad\frac{G_{55}}{\alpha'}=\frac{2}{\ell^2}\,c^{-\frac{1}{2}}k q\,(2\nu_1\nu_2\nu_3)^{-\frac{1}{2}}\nu_1\nu_2&\\\nonumber
&G_{{\sss \mathrm{O}}}=4k^4q^3c^{-1}\nu_1\nu_2\nu_3\,\vert ab-4\nu_2\nu_3\vert^{\frac{1}{2}}&
\ee
One might worry about potential divergencies and zeros caused by the factor involving $a$ and $b$, but writing this in terms of the $\nu$'s we get
\bq
ab-4\nu_2\nu_3=(1+\nu_1)^2-(\nu_2+\nu_3)^2
\eq
which is strictly positive in the genuine rank 6 case. The theory with the above open string data is also an NCOS, since $\Theta^{01}\neq 0$ and the effective open string tension is finite. The strong and weak coupling limits of ordinary NCOS, obtained in the (1,0) case, corresponds to letting $k$ go to zero and infinity, respectively. Since the effective coupling for the open $(p,q)$-string in question scales with a positive power of $k$, we see that the NCOS theory defined by this string is S-dual to ordinary NCOS.

Turning to the case ${\cal Q}\neq 0$, corresponding to arbitrary values of the constant part of the axion, and charges such that $p-q\tilde{q}\neq 0$, $\Theta^{01}$ scales like $\epsilon$ and therefore goes to zero in the UV limit. The space-space noncommutativity parameters on the other hand are finite (and are just obtained from (\ref{rank6})). The first block of the open string metric diverges whereas the other two are finite. Hence the effective open string tension diverges. The open string coupling scales like $\epsilon^{-1/2}$ and therefore diverges in the UV. We are thus lead to believe that the $(p,q)$-strings describe a noncommutative Yang-Mills theory in this case. For a general D$p$-brane in $B$-fields, the coupling of NCYM is \cite{Seiberg:1999}
\bq\label{gym}
g_{{\sss \mathrm{YM}}}^2=G_{{\sss \mathrm{O}}}(\alpha')^{\frac{p-3}{2}}\sim G_{{\sss \mathrm{O}}}\epsilon^{\frac{p-3}{4}}
\eq
where we have used the usual scaling of $\alpha'$. Thus we see that we get a finite coupling for the NCYM theory above. This theory should only be considered as an effective low energy theory, since it is not UV-complete. We, however, get the peculiarity that the open string metric only blows up in the electric directions without noncommutativity, and in this sense the theory differs from ordinary NCYM.

The open string data of the (1,0) NCOS do not depend on the axion, but as we have seen, this is not true for the SL(2,$\ZZ$)-transformed theory. For rational $\tilde{q}$ and charges fulfilling $p-q\tilde{q}=0$  we get an S-dual NCOS obtained from $(p,q)$-strings. For arbitrary $\tilde{q}$ we can choose charges such that we get NCYM which is, however, not UV-complete.

\subsubsection{Electric rank 4}
Now consider the electric rank 4 case with $\nu_3=0$. The
background corresponds to an (F,D1,D3,D5) bound state, with the
D3-brane along $x^0,x^1,x^4,x^5$. Define \bq f_4={\cal
Q}^2\Delta_{{\sss +-}}+k^4 q^2\Delta_{{\sss --}}\Delta_{{\sss ++}}
\eq In the fixed coordinates, the open string data for arbitrary
charges of the strings on the probe are \be\nonumber
\Theta^{01}&=&-\ell^2k\,\sqrt{2\nu_1}\,{\cal Q}\,({\cal
Q}^2+k^4q^2\Delta_{{\sss ++}})^{-1}\\\nonumber
\Theta^{23}&=&-\ell^2k\,\sqrt{2\nu_2}\,{\cal Q}\,({\cal
Q}^2+k^4q^2\Delta_{{\sss --}})^{-1}\\\nonumber
\Theta^{45}&=&-2\ell^2 k^3 q\,\sqrt{2\nu_1\nu_2}\,\Delta_{{\sss +-}}(f_4+4k^4q^2\nu_1\nu_2)^{-1}\\
{\scs \frac{G_{11}}{\alpha'}}&=&{\scs \frac{1}{\ell^2}}\,k^{-1}f_4^{-\frac{1}{2}}({\cal Q}^2+k^4q^2\Delta_{{\sss ++}})\\\nonumber
{\scs \frac{G_{33}}{\alpha'}}&=&{\scs \frac{1}{\ell^2}}\,k^{-1}f_4^{-\frac{1}{2}}({\cal Q}^2+k^4q^2\Delta_{{\sss --}})\\\nonumber
{\scs \frac{G_{55}}{\alpha'}}&=&{\scs \frac{1}{\ell^2}}\,k^{-1}f_4^{-\frac{1}{2}}(f_4+4k^4q^2\nu_1\nu_2)\\\nonumber
G_{{\sss \mathrm{O}}}&=&k^{-2}\Delta_{{\sss +-}}^{-\frac{1}{2}}f_4^{-\frac{1}{2}}({\cal Q}^2+k^4q^2\Delta_{{\sss ++}})^{\frac{1}{2}}({\cal Q}^2+k^4q^2\Delta_{{\sss --}})^{\frac{1}{2}}(f_4+4 k^4 q^2\nu_1\nu_2)^{\frac{1}{2}}
\ee
Now consider the UV limit
\bq
\Delta_{{\sss --}}\sim \epsilon\,,\quad\Delta_{{\sss ++}}\sim 1+\nu_1+\nu_2\equiv a\,,\quad \Delta_{{\sss +-}}\sim 1+\nu_1-\nu_2\equiv b
\eq
For the charges (1,0) we then get
\be\nonumber
&\Theta^{01}=-\ell^2k\,\sqrt{2\nu_1}\,,\quad \Theta^{23}=-\ell^2k\,\sqrt{2\nu_2}\,,\quad \Theta^{45}=0&\\
&{\scs \frac{G_{\mu\nu}}{\alpha'}}={\scs \frac{1}{\ell^2}}\,k^{-1}b^{-\frac{1}{2}}\eta_{\mu\nu}\,,\quad G_{{\sss \mathrm{O}}}=k^{-2}b^{-\frac{1}{2}}&
\ee
We see that we get space-time as well as space-space noncommutativity, and that the effective open string tension is finite, and hence we get an NCOS as expected.

In the rank 4 case we also have the possibility of having SL(2,$\ZZ$)-duals, since \eg D-strings can end on the D3-brane in the bound state. For the rank 4 and 2 cases, the axion in the D5-brane solution is constant, $\chi=\tilde{q}$. In the case of $(p,q)$-strings with ${\cal Q}=0$, which can be obtained for rational axion, we get
\be\nonumber
&\Theta^{01}=\Theta^{23}=0\,,\quad \Theta^{45}=-\frac{1}{2}\,\ell^2k^{-1}q^{-1}b(\nu_1\nu_2)^{-\frac{1}{2}}&\\\nonumber
&{\scs \frac{G_{11}}{\alpha'}}={\scs \frac{1}{\ell^2}}k^{-1}qa^{\frac{1}{2}}\epsilon^{-\frac{1}{2}}\,,\quad{\scs \frac{G_{33}}{\alpha'}}={\scs \frac{1}{\ell^2}}k^{-1}qa^{\frac{1}{2}}\epsilon^{\frac{1}{2}}\,,\quad{\scs \frac{G_{55}}{\alpha'}}={\scs \frac{4}{\ell^2}}k^{-1}qa^{\frac{1}{2}}b^{-1}\nu_1\nu_2\epsilon^{-\frac{1}{2}}&\\
&G_{{\sss \mathrm{O}}}=2k^2 q^2 b^{-\h}\sqrt{\nu_1\nu_2}&
\ee
Thus we only get space-space noncommutativity and we also see that the effective open string tension diverges, so the resulting theory appears to be a NCYM. Regarded as a theory on the D5-brane, the Yang-Mills coupling goes to zero in the UV, yielding a free theory. However, it is seen that the open string metric blows up on the D3-brane, in agreement with the fact that the dual strings only can end on the D3-brane. The Yang-Mills coupling on the D3-brane equals the open string coupling, so it seems reasonable that we get a well-defined NCYM on the D3-brane. From the $k$-scaling of the couplings we see that we get a strong/weak coupling relation between this NCYM theory and NCOS, but it is not clear how this should be understood, since we are dealing with four- and six-dimensional theories.

For arbitrary axion and charges $(p,q)$ for the strings on the probe, such that ${\cal Q}\neq 0$, the open string data become
\be\nonumber
&\Theta^{01}=-\ell^2k\,\sqrt{2\nu_1}\,{\cal Q}\,({\cal Q}^2+k^4q^2a)^{-1}\,,\quad
\Theta^{23}=-\ell^2k\,\sqrt{2\nu_2}\,{\cal Q}^{-1}&\\\nonumber
&\Theta^{45}=-2 \ell^2k^3 q\,\sqrt{\nu_1\nu_2}\,b\,({\cal Q}^2 b+4k^4q^2\nu_1\nu_2)^{-1}\,,\quad{\scs \frac{G_{11}}{\alpha'}}=\frac{1}{{\cal Q}\ell^2}\,k^{-1}b^{-\h}({\cal Q}^2+k^4q^2a)&\\\nonumber
&{\scs \frac{G_{33}}{\alpha'}}=\frac{{\cal Q}}{\ell^2}\,k^{-1}b^{-\h}\,,\quad
{\scs \frac{G_{55}}{\alpha'}}=\frac{1}{{\cal Q}\ell^2}\,k^{-1}b^{-\frac{3}{2}}({\cal Q}^2b+4k^4q^2\nu_1\nu_2)&\\
&G_{{\sss \mathrm{O}}}=k^{-2}b^{-1}({\cal Q}^2+k^4q^2a)^{\h}({\cal Q}^2b+4k^4q^2\nu_1\nu_2)^{\h}&
\ee
Thus we both get space-time and space-space noncommutativity. Furthermore, the effective open string tension is finite, so we get an NCOS. In general this NCOS is not S-dual to the ordinary NCOS obtained from the (1,0) charges, since both couplings diverge when $k$ goes to zero. As for the D3-brane, we have a special case, for certain values of the charges. To be specific, the charges should be such that ${\cal Q}$ scales like $q k^\beta$, with $\beta>1$. The coupling of the $(p,q)$ NCOS  then scales with a positive power of $k$, yielding a weakly coupled theory when the (1,0) NCOS is strongly coupled and vice versa. Hence the two NCOS theories are S-dual.

We have seen that for arbitrary axion we can choose charges such that (1,0) NCOS is S-dual to $(p,q)$ NCOS. In the special case $p-q\chi=0$, which can be obtained for rational axion, the S-dual seems to be NCYM on the D3-brane sitting in the D5-brane.

\subsubsection{Magnetic rank 4}
This case corresponds to $\nu_1=0$ and the background is then a (D1,D3,D3,D5) bound state, where the D3-branes share the electric directions as in the rank 6 case. Define
\bq
h_4={\cal Q}^2\Delta_{{\sss --}}+k^4q^2\Delta_{{\sss +-}}\Delta_{{\sss -+}}
\eq
In the fixed coordinates, the open string data then become
\be\nonumber
\Theta^{01}&=&2\ell^2k^3 q\sqrt{\nu_2\nu_3}\,\Delta_{{\sss --}}(h_4-4k^4q^2\nu_2\nu_3)^{-1}\\\nonumber
\Theta^{23}&=&-\ell^2k\,\sqrt{2\nu_2}\,{\cal Q}\,({\cal Q}^2+k^4q^2\Delta_{{\sss -+}})^{-1}\\\nonumber
\Theta^{45}&=&-\ell^2k\,\sqrt{2\nu_3}\,{\cal Q}\,({\cal Q}^2+k^4q^2\Delta_{{\sss +-}})^{-1}\\
{\scs \frac{G_{11}}{\alpha'}}&=&{\scs \frac{1}{\ell^2}}\,k^{-1}h_4^{-\frac{1}{2}}(h_4-4k^4q^2\nu_2\nu_3)\\\nonumber
{\scs \frac{G_{33}}{\alpha'}}&=&{\scs \frac{1}{\ell^2}}\,k^{-1}h_4^{-\frac{1}{2}}({\cal Q}^2+k^4q^2\Delta_{{\sss -+}})\\\nonumber
{\scs \frac{G_{55}}{\alpha'}}&=&{\scs \frac{1}{\ell^2}}\,k^{-1}h_4^{-\frac{1}{2}}({\cal Q}^2+k^4q^2\Delta_{{\sss +-}})\\\nonumber
G_{{\sss \mathrm{O}}}&=&k^{-2}\Delta_{{\sss --}}^{-\frac{1}{2}}h_4^{-\frac{1}{2}}({\cal Q}^2+k^4q^2\Delta_{{\sss -+}})^{\frac{1}{2}}({\cal Q}^2+k^4q^2\Delta_{{\sss +-}})^{\frac{1}{2}}(h_4-4 k^4 q^2\nu_2\nu_3)^{\frac{1}{2}}
\ee
The open string data are then analysed in the UV limit
\bq
\Delta_{{\sss --}}\sim\epsilon\,,\quad \Delta_{{\sss +-}}\sim 1+\nu_2-\nu_3\equiv a\,,\quad \Delta_{{\sss -+}}\sim 1-\nu_2+\nu_3\equiv b
\eq
For the (1,0)-string we get
\be
&\Theta^{01}=0\,,\quad \Theta^{23}=-\ell^2k\,\sqrt{2\nu_2}\,,\quad \Theta^{45}=-\ell^2k\,\sqrt{2\nu_3}&\\
&{\scs \frac{G_{\mu\nu}}{\alpha'}}={\scs \frac{1}{\ell^2}}\epsilon^{-\frac{1}{2}}\,k^{-1}\eta_{\mu\nu}\,,\quad G_{{\sss \mathrm{O}}}=k^{-2}\epsilon^{-\frac{1}{2}}&
\ee
We get space-space noncommutativity and a diverging effective open string tension, suggesting a NCYM. It is seen that the open string metric blows up on the entire D5-brane. We also see that we get a finite Yang-Mills coupling  on the D5-brane and therefore we get an effective NCYM theory obtained from the (1,0) strings.

Now we look at the SL(2,$\ZZ$)-transformed $(p,q)$-strings. Starting with rational axion and ${\cal Q}=0$, the open string data are
\be\nonumber
&\Theta^{01}=\ell^2k^{-1}q^{-1}\sqrt{\nu_2\nu_3}\,,\quad\Theta^{23}=\Theta^{45}=0&\\\nonumber
&{\scs \frac{G_{11}}{\alpha'}}={\scs \frac{2}{\ell^2}}\,k\,q(ab)^{-\frac{1}{2}}\,,\quad{\scs \frac{G_{33}}{\alpha'}}={\scs \frac{1}{\ell^2}}\,k\,qb(ab)^{-\frac{1}{2}}\,,\quad{\scs \frac{G_{55}}{\alpha'}}={\scs \frac{1}{\ell^2}}\,k\,qa(ab)^{-\frac{1}{2}}&\\
&G_{{\sss \mathrm{O}}}=\sqrt{2}k^2 q^2&
\ee
where we have the used the following UV-behaviour for a term entering many of the quantities
\bq
\Delta_{{\sss +-}}\Delta_{{\sss -+}}-4\nu_2\nu_3=1-(\nu_2+\nu_3)^2+2\frac{R^2}{u^2}+\frac{R^4}{u^4}\sim 2\epsilon
\eq
We only get space-time noncommutativity. Furthermore, the effective open string tension is finite so we get an NCOS. From the scaling of the coupling with respect to $k$, it is seen that this NCOS is S-dual to the NCYM above.

For the $(p,q)$-string with arbitrary axion and ${\cal Q}\neq 0$, we get
\be\nonumber
&\Theta^{01}=2\ell^2k^3q\sqrt{\nu_2\nu_3}\,({\cal Q}^2+2k^4q^2)^{-1}\,,\quad
\Theta^{23}=-\ell^2k\,\sqrt{2\nu_2}\,{\cal Q}\,({\cal Q}^2+k^4q^2b)^{-1}&\\\nonumber
&\Theta^{45}=-\ell^2k\,\sqrt{2\nu_3}\,{\cal Q}\,({\cal Q}^2 b+k^4q^2a)^{-1}\,,\quad{\scs \frac{G_{11}}{\alpha'}}=\frac{1}{\ell^2}k^{-3}q^{-1}(ab)^{-\h}({\cal Q}^2+2k^4q^2)&\\\nonumber
&{\scs \frac{G_{33}}{\alpha'}}=\frac{1}{\ell^2}k^{-3}q^{-1}(ab)^{-\h}({\cal Q}^2+k^4q^2b)\,,\quad
{\scs \frac{G_{55}}{\alpha'}}=\frac{1}{\ell^2}k^{-3}q^{-1}(ab)^{-\frac{1}{2}}({\cal Q}^2+k^4q^2a)&\\
&G_{{\sss \mathrm{O}}}=\sqrt{2}k^{-2}(ab)^{-\h}({\cal Q}^2+k^4q^2a)^{\h}({\cal Q}^2+k^4q^2b)^{\h}&
\ee
We get space-time as well as space-space noncommutativity and the effective open string tension is finite so we get an NCOS. In general this NCOS is not S-dual to the NCYM obtained with (1,0)-strings, since both couplings diverge when $k$ goes to zero. But again, having charges yielding a ${\cal Q}$ which scales like $q k^\beta$ with $\beta>1$, the NCOS coupling will scale with a positive power of $k$ and the two theories are therefore S-dual.

\subsubsection{Electric rank 2}
This case corresponds to $\nu_2=\nu_3=0$. The background is an (F,D1,D5) bound state and this is the only case where there is no D3-brane present in the bound state. We might therefore run into problems when using general strings to describe the theory on the brane. Take a (D1,NS5) bound state as example. The F-string cannot end on the NS5-brane, but it can end on the D-string sitting inside the NS5-brane. A general $(p,q)$-string, with $q\neq 0$, may however end on the NS5-brane. By SL(2,$\ZZ$)-covariance, a D-string may end on the F-string in an (F,D5) bound state. When the string charges of the string in the bound state and the string on the probe are opposite we therefore get the above kinematical constraint.  SL(2,$\ZZ$)-covariantising this statement gives the result that when  $\epsilon_{rs}p^{r}\hat{p}^s=p^r\hat{p}_r=0$, where $\hat{p}_r$ are the string charges in the bound state, the $(p,q)$-strings can only end on the strings in the 5-brane. Otherwise, the strings may end on the entire 5-brane. Since the charges of the strings in the bound state can be obtained from the 2-forms of the brane solution, we can be certain that a $(p,q)$-string only can end on the strings in the 5-brane if the following relation holds
\be\nonumber
p^r C_{r}=0
\ee
As seen from the D5-brane solution in the electric rank 2 case, this corresponds to $p-q\chi={\cal Q}=0$.

In fixed coordinates, the open string data are
\be\nonumber
&\Theta^{01}=-\ell^2k\,\sqrt{2\nu_1}\,{\cal Q}\,({\cal Q}^2+k^4q^2\Delta_{{\sss +}})^{-1}\,,\Theta^{23}=\Theta^{45}=0&\\\nonumber
&\frac{G_{11}}{\alpha'}=\frac{1}{\ell^2}\,k^{-1}\,\Delta_+^{-\h}({\cal Q}^2+k^4q^2\Delta_{{\sss +}})({\cal Q}^2+k^4q^2\Delta_{{\sss -}})^{-\h}&\\
&\frac{G_{33}}{\alpha'}=\frac{G_{55}}{\alpha'}=\frac{1}{\ell^2}\,k^{-1}\,\Delta_{{\sss +}}^{-\h}({\cal Q}^2+k^4q^2\Delta_{{\sss +}})^{\h}&\\\nonumber
&G_{{\sss \mathrm{O}}}=k^{-2}\Delta_{{\sss +}}^{-\h}({\cal Q}^2+k^4q^2\Delta_{{\sss -}})^{\h}({\cal Q}^2+k^4q^2\Delta_{{\sss +}})^{\h}&
\ee
In this case, the UV limit is
\bq
\Delta_{{\sss -}}\sim \epsilon\,,\quad \Delta_{{\sss +}}\sim 2
\eq
 For (1,0)-strings the open string data become
\be\nonumber
&\Theta^{01}=-\ell^2k\,\sqrt{2}\,,\quad\Theta^{23}=\Theta^{45}=0&\\
&{\scs \frac{G_{\mu\nu}}{\alpha'}}={\scs \frac{1}{\sqrt{2}\ell^2}}\,k^{-1}\,\eta_{\mu\nu}\,,\quad G_{{\sss \mathrm{O}}}={\scs \frac{1}{\sqrt{2}}}\,k^{-2}&
\ee
With space-time noncommutativity and finite effective open string tension we get an NCOS as expected.

Consider the case ${\cal Q}=0$. Then the open strings can only end on the strings sitting in the 5-brane. In this case, the open string data are
\be\nonumber
&\Theta^{01}=\Theta^{23}=\Theta^{45}=0&\\\nonumber
&\frac{G_{11}}{\alpha'}=\frac{1}{\ell^2}\,k\,q\,\epsilon^{-\h}\,,\quad
\frac{G_{33}}{\alpha'}=\frac{G_{55}}{\alpha'}=\frac{1}{\ell^2}\,k\,q\,\epsilon^{\h}&\\
&G_{{\sss \mathrm{O}}}=\sqrt{2}\,k^{2}q^2\epsilon^{\h}&
\ee
The open string metric blows up in the electric directions and shrinks in the magnetic directions. From the scaling of the coupling and (\ref{gym}), we see that we get a finite Yang-Mills coupling in two dimensions. All noncommutativity parameters vanish and we thus get an effective description in terms of a 2d YM theory on the strings in the 5-brane.

If we want to look for a six-dimensional S-dual of the NCOS, we therefore have to consider the case ${\cal Q}\neq 0$, and then the open string data are
\be\nonumber
&\Theta^{01}=-\ell^2k\,\sqrt{2}\,{\cal Q}\,({\cal Q}^2+2k^4q^2)^{-1}\,,\Theta^{23}=\Theta^{45}=0&\\\nonumber
&\frac{G_{11}}{\alpha'}=\frac{1}{\sqrt{2}\ell^2{\cal Q}}\,k^{-1}\,({\cal Q}^2+2k^4q^2)\,,\quad
\frac{G_{33}}{\alpha'}=\frac{G_{55}}{\alpha'}=\frac{{\cal Q}}{\sqrt{2}\ell^2}\,k^{-1}&\\
&G_{{\sss \mathrm{O}}}={\scs \frac{{\cal Q}}{\sqrt{2}}}\,k^{-2}({\cal Q}^2+2k^4q^2)^{\h}&
\ee
We get space-time noncommutativity and finite effective open string tension. Even in this case we get an S-dual NCOS when the charges have values such that ${\cal Q}$ scales like $q k^\beta$, with $\beta>1$. In particular, when $\beta=2$, the open string data take the simple form
\be
&\Theta^{01}=-\ell^23\sqrt{2}\,k^{-1}\,q^{-2}\,,\Theta^{23}=\Theta^{45}=0&\\\nonumber
&\frac{G_{11}}{\alpha'}=\frac{3}{\sqrt{2}\ell^2}\,k\,q^2\,,\quad
\frac{G_{33}}{\alpha'}=\frac{G_{55}}{\alpha'}=\frac{1}{\sqrt{2}\ell^2}\,k&\\
&G_{{\sss \mathrm{O}}}={\scs \sqrt{\frac{3}{2}}}\,k^{2}q^2&
\ee

\subsubsection{Magnetic rank 2}
This case corresponds to $\nu_1=\nu_2=0$. The background is a (D3,D5) bound state.
In fixed coordinates the open string data are
\be\nonumber
&\Theta^{45}=-\ell^2k\,\sqrt{2\nu_3}\,{\cal Q}\,({\cal Q}^2+k^4q^2\Delta_{{\sss -}})^{-1}\,,\Theta^{01}=\Theta^{23}=0&\\\nonumber
&\frac{G_{11}}{\alpha'}=\frac{G_{33}}{\alpha'}=\frac{1}{\ell^2}\,k^{-1}\Delta_{{\sss -}}^{-\h}({\cal Q}^2+k^4q^2\Delta_{{\sss +}})^{\h}&\\\nonumber
&\frac{G_{55}}{\alpha'}=\frac{1}{\ell^2}\,k^{-1}\Delta_{{\sss -}}^{-\h}({\cal Q}^2+k^4q^2\Delta_{{\sss -}})({\cal Q}^2+k^4q^2\Delta_{{\sss +}})^{-\h}&\\
&G_{{\sss \mathrm{O}}}=k^{-2}\Delta_{{\sss -}}^{-\h}({\cal Q}^2+k^4q^2\Delta_{{\sss -}})^{\h}({\cal Q}^2+k^4q^2\Delta_{{\sss +}})^{\h}&
\ee
The UV limit is the same as in the electric rank 2 case. For a (1,0)-string we get
\be\nonumber
&\Theta^{45}=-\ell^2k\,\sqrt{2}\,,\quad\Theta^{01}=\Theta^{23}=0&\\
&{\scs \frac{G_{\mu\nu}}{\alpha'}}={\scs \frac{1}{\ell^2}}\,k^{-1}\epsilon^{-\h}\,\eta_{\mu\nu}\,,\quad G_{{\sss \mathrm{O}}}=k^{-2}\epsilon^{-\h}&
\ee
We only have space-space noncommutativity and the effective open string tension diverges. From the scaling of $G_{{\sss \mathrm{O}}}$, it is seen that we get a finite Yang-Mills coupling. We therefore have an effective description in terms of NCYM.
For $(p,q)$-strings with rational axion and ${\cal Q}=0$, the open string data become
\be\nonumber
&\Theta^{01}=\Theta^{23}=\Theta^{45}=0&\\\nonumber
&{\scs \frac{G_{11}}{\alpha'}}={\scs \frac{G_{33}}{\alpha'}}={\scs \frac{\sqrt{2}}{\ell^2}}\,k\,q\,\epsilon^{-\h}\,,\quad {\scs \frac{G_{55}}{\alpha'}}={\scs \frac{1}{\sqrt{2}\ell^2}}\,k\,q\,\epsilon^{\h}&\\
&G_{{\sss \mathrm{O}}}=\sqrt{2}k^{2}q^2&
\ee
All $\Theta$ parameters vanish, the effective open string tension diverges, the open string metric only blows up on the D3-brane and the open string coupling is finite so the result looks like an ordinary Yang-Mills theory on the D3-brane sitting in the D5-brane.

On the other hand, for arbitrary axion and ${\cal Q}\neq 0$, we get
\be\nonumber
&\Theta^{45}=-\ell^2k\,\sqrt{2\nu_3}\,{\cal Q}^{-1}\,,\quad\Theta^{01}=\Theta^{23}=0&\\\nonumber
&{\scs \frac{G_{11}}{\alpha'}}={\scs \frac{G_{33}}{\alpha'}}={\scs \frac{1}{\ell^2}}\,k^{-1}({\cal Q}^2+2k^4q^2)^{\h}\epsilon^{-\h}\,,\quad {\scs \frac{G_{55}}{\alpha'}}={\scs \frac{{\cal Q}^2}{\ell^2}}\,k^{-1}({\cal Q}^2+2k^4q^2)^{-\h}\epsilon^{-\h}&\\
&G_{{\sss \mathrm{O}}}=k^{-2}{\cal Q}\,({\cal Q}^2+2k^4q^2)^{\h}\epsilon^{-\h}&
\ee
We get space-space noncommutativity, a diverging effective open string tension and finite Yang-Mills coupling. The resulting theory is therefore a NCYM on the D5-brane.

In the magnetic rank 2 case, open D3-branes should be light, due to the presence of the critical electric 4-form potential. The proper description of the theory on the 5-brane is therefore not in terms of the open strings but instead the open D3-branes, yielding the OD3 theory.

\section{Discussion}

In this paper we have constructed an SL(2,$\ZZ$)-covariant
generalisation of NCOS and NCYM on the D3-brane and NCOS 
on the D5-brane in type IIB. In particular, we have obtained
SL(2,$\ZZ$)-covariant expressions for the effective open string
metric, the noncommutativity parameter and the effective open
string coupling. We then inserted the fields of the relevant
D3-brane and D5-brane supergravity backgrounds in these
expressions, and the decoupled theories on the brane are obtained from a unique UV limit. As a
result we get noncommutative open $(p,q)$-string theories, with
the string charges appearing explicitly in the open string data.
As mentioned, it is the relative angle between the background
2-form doublet and the doublet of charges of the strings on the
probe which determines what kind of theory we get. By transforming either the background or the open strings ending on the probe, we get new theories, and in this way we can examine the duality properties of the theories. We get agreement with previous results for the
D3-brane \cite{Russo:2000}, namely for rational axion, NCOS is
S-dual to a NCYM. Previously, this was obtained by just transforming the background, whereas we just transform the open strings ending on the probe. The NCYM is then obtained from a string with charges fulfilling $p-\chi q=0$. We believe that this result yields strong evidence for the validity of our method.

For arbitrary axion and charges such that
$p-q\chi\neq 0$, we always get an NCOS, which we call $(p,q)$ NCOS, since the theory contains light $(p,q)$-strings. For certain charges
$(p,q)$, this NCOS is S-dual to ordinary NCOS. As mentioned, it is crucial to keep the background fixed in this process, since transforming the background as well to get both electric NS-NS and RR 2-forms, yields a theory equivalent (instead of S-dual) to NCOS. We have seen that in a generic case, the S-dual of ordinary NCOS is a $(p,q)$ NCOS and only in certain special cases do we get NCYM. 

Similarly, on the D5-brane The ordinary
NCOS is obtained in the electric rank 2 case. We can
get an S-dual NCOS for arbitrary values of the axion and charges
such that $p-q\tilde{q}\neq 0$. A property which holds for most of
the NCOS theories considered in this paper. All other 2-form
configurations were also analysed. In the rank 6 case, a (1,0)
NCOS with rational constant part of the axion is S-dual to a $(p,q)$ NCOS when
$p-q\tilde{q}=0$. For arbitrary axion and $p-q\tilde{q}\neq0$, the
dual theory has an effective description in terms of NCYM. In the
electric rank 4 case with rational axion, (1,0) NCOS has an
SL(2,$\ZZ$)-dual description in terms of NCYM on the D3-brane in
the D5-brane. This theory is obtained from strings with charges
fulfilling $p-q\chi=0$. In the magnetic rank 4 case, NCYM is
S-dual to NCOS for arbitrary values of the axion. In the magnetic
rank 2 case, we only have NCYM descriptions. In the latter case,
the strings are not the proper objects to describe the theory on
the brane, since they are not the lightest objects. This role is
instead played by D3-branes. The NCYM theories obtained in the
magnetic rank 2 case are therefore only effective descriptions of
the OD3-theory. A proper description of this theory would require
quantising the D3-branes, something which is not yet possible. A
first step could instead be to generalise the open string data to
the case of branes. 

When we have a critical field, the asymptotic closed string metric
gives a hint as to which objects become light. In the magnetic
rank 2 case, the metric corresponds to a smeared 3-brane
\cite{Cederwall:1999}. In this case we get light D3-branes and the
proper theory is OD3. Actually, this is the only case where the
asymptotic closed string metric corresponds to a smeared 
3-brane. In all other cases, the metric is that of a smeared 
string. This is in agreement with the fact that we can get light
strings in these cases. According to this, there should not be
light D3-branes in the higher rank cases, a statement which is confirmed by the behaviour of the 4-form. In the critical field
limit, the 4-form diverges logarithmically in the UV
\cite{Gran:2001:1}, and therefore we cannot get a
cancellation of the divergence of the usual D3-brane tension. 

It also follows from the above that the deformed 5-brane solution cannot describe the OD5 theory, since we never get an asymptotic closed string metric which blows up on the entire 5-brane. The only way to get SO(1,5) isometry on the 5-brane is by setting all the deformation parameters $\nu_i$ to zero. Indeed, it can be shown that the NS5 supergravity solution with a 6-form RR-potential on the brane, which can be obtained by a double T-duality on the (NS5,D3) bound state, corresponds to an ordinary $(p,q)$ 5-brane\footnote{This was also noted in \cite{Alishahiha:2000:2}.}. The 6-form is just the potential for the 7-form which is Hodge dual to the 3-form giving the D5-brane charge. The critical 6-form is obtained by choosing the integration constant in the harmonic function to be zero. By doing this, one also ends up with an asymptotic metric which instead of becoming minkowskian blows up on the entire 5-brane. It can also be seen from the supergravity solution that we get a finite effective tension for the D5-branes which therefore become light. Similarly, for the ordinary undeformed D$p$-brane solutions, the D$p$-branes themselves become light due to the compensating electric $(p+1)$-form potential, which becomes critical in the UV if we choose the constant in the harmonic function to be zero.  By doing this, the full metric becomes $\mathrm{AdS}_{p+2}\times S^{8-p}$, and therefore the asymptotic metric blows up on the entire D$p$-brane. It is not clear to us, though, how these light branes should be interpreted.

Recently, ideas related to our open string theories have been discussed in
\cite{Lu:2001}, from the world volume point of view, and in \cite{Larsson:2001},
from the supergravity duals. In these papers new theories of open D-branes are
proposed, \eg, on the D3-brane a theory of open D-strings, called $\widetilde{\mathrm{OD}1}$ in \cite{Larsson:2001}, is proposed to be the strong
coupling limit of NCOS. This case involves a vanishing axion and should
therefore be compared to the usual duality between NCOS and NCYM. In the latter
paper, the D-strings are argued to be the solitons of NCYM,
which in the limit of large noncommutativity parameter becomes light. Obtaining
the $\widetilde{\mathrm{OD}1}$ theory from NCYM thus involves an additional limit. From this
picture it is clear that the D-strings of $\widetilde{\mathrm{OD}1}$ theory are strongly coupled
when viewed as fundamental objects, but interact weakly through exchanging
weakly coupled F-strings (with large tension). By taking $G_{{\sss \mathrm{O}}}$
large in the open string data for NCOS in (\ref{NCOSdata}), \ie, taking $c$
large, and also rescaling the coordinates in order to get light
D-strings, we can get the open
string data for $\widetilde{\mathrm{OD}1}$\footnote{Note that studying
F-string theory in an (F,D3) background is equivalent to studying D-string
theory in an (D1,D3) background. This implies that perturbative D-strings in an
electric $C$-field is equivalent to NCOS.}, but since the D-string is
strongly coupled and there exist weakly coupled objects (the F-strings) the
D-string is not a proper fundamental object, it is solitonic.
In our picture, we should therefore instead study the theory on the probe brane using the perturbative F-strings, leading to
NCYM in which the D-string appears as a soliton.
There is therefore no contradiction between the
$\widetilde{\mathrm{OD}1}$ theory and our $(p,q)$-string theories since the open
$(p,q)$-string data are only relevant when the $(p,q)$-string is weakly coupled. On the 5-branes it seems reasonable that the OD$p$ theories are the proper descriptions instead of NCYM. However, in the cases where $p-q\chi\neq 0$ (as well as $p-q\chi=0$ for rank 6) we can have a well defined $(p,q)$ NCOS on the 5-brane, and in these cases no results have been derived for the OD$p$-theories.

There is another important aspect which should be mentioned. It is known that little strings live on the 5-branes \cite{Seiberg:1997:2}, and that these will dominate in certain regimes of the noncommutative theories, at least in the rank 2 case \cite{Harmark:2000:2}. One could speculate that the D-strings in \cite{Larsson:2001} play the same role on the D3-brane as the little strings on the 5-branes, since both strings can be regarded as solitons of the NCYM on the brane. 
In the present work the little strings have not been considered. An analysis including the little strings might modify some of the conclusions reached in this paper.

\vspace{5mm}
{\Large {\bf Acknowledgements}}
\vspace{5mm}\\
We would like to thank M. Cederwall, B.E.W. Nilsson, H. Larsson, V. Lima Campos, D. Berman and P. Sundell for valuable discussions. This work is partly supported by the Danish Natural Science Research Council.

\appendix

\section{Type IIB supergravity}

Type IIB supergravity in ten dimensions has an SL(2,$\RR$) invariance (which is broken to SL(2,$\ZZ$) by quantum effects) and contains the following fields: the metric, 2 scalars (the dilaton $\phi$ and the axion $\chi$), the NS-NS 2-form potential $B$, the R-R 2-form potential $C$ and the R-R 4-form potential $C_{(4)}$. There exists a formulation with manifest SL(2,$\RR$) covariance \cite{Schwarz:1983,Howe:1984}. Here we use the notation of \cite{Cederwall:1997:1,Cederwall:1997:2}. The two 2-forms can be collected in an SL(2,$\RR$) doublet $C_{r}$, where $r$=1,2 corresponds to the NS-NS and R-R 2-forms respectively. The scalars can be described by a complex doublet ${\cal U}^r$, with $\tau={\cal U}^1/{\cal U}^2=\chi+i\, e^{-\phi}$. The scalar doublet fulfills the SL(2,$\ZZ$)-invariant constraint
\bq
{\scs \frac{i}{2}}\epsilon_{rs}\,{\cal U}^r\bar{{\cal U}}^s=1
\eq
The 2-form doublet has a 3-form doublet of field strengths $H_{(3)r}=dC_{r}$, which can be combined with the scalar doublet into a complex 3-form
\bd
{\cal H}_{(3)}={\cal U}^r H_{(3)r}\,,\qquad H_{(3)r}=\epsilon_{rs}\,\mathrm{Im}\big(\,{\cal U}^s\bar{{\cal H}}_{(3)}\big)
\ed
From the scalar doublet we can construct the Mauer-Cartan 1-forms $P$ and $Q$
\bq
Q={\scs \frac{1}{2}}\,\epsilon_{rs}d{\cal U}^r\bar{{\cal U}}^s\,,\qquad
P={\scs \frac{1}{2}}\,\epsilon_{rs}d{\cal U}^r{\cal U}^s
\eq

The equations of motion can now be written as
\be\nonumber
&&D{\ast}P+{\scs \frac{i}{4}}\,{\cal H}_3\wdg\ast{\cal H}_3=0\\\nonumber
&&D{\ast}{\cal H}_3+i\,P\wdg\ast\bar{\cal H}_3-i\,H_5\wdg{\cal H}_3=0\\
&&D{\cal H}_3+i\,\bar{\cal H}_3\wdg P=0\\\nonumber &&dH_5-{\scs
\frac{i}{2}}\,{\cal H}_3\wdg\bar{\cal H}_3=0\\\nonumber &&R_{{\sss
MN}}=2 \bar{P}_{{\sss (M}} P_{{\sss N)}}+{\scs
\frac{1}{4}}\bar{\cal H}_{{\sss (M}}{}^{{\sss RS}}{\cal H}_{{\sss
N)RS}}-{\scs \frac{1}{48}}\,g_{{\sss MN}}\bar{\cal H}_{{\sss
RST}}{\cal H}^{{\sss RST}}+{\scs \frac{1}{96}}\,H_{{\sss
(M}}{}^{{\sss RSTU}}H_{{\sss N)RSTU}} \ee The first two equations
are the equations of motion for $P$ and ${\cal H}_3$,
respectively. The following two are the Bianchi identities for the
3-forms and the 5-form. The last line is the Einstein equations.

%
%

\bibliographystyle{JHEP}
\bibliography{pqjhep}

\end{document}